\documentclass[preprint,12pt]{elsarticle}
\usepackage{graphicx}
\usepackage{mathptmx}      

\usepackage{amssymb}
\usepackage{amsmath,amsthm}
\usepackage[colorlinks=true,allcolors=blue]{hyperref}
\usepackage{cleveref}
\usepackage[subnum]{cases}
\usepackage{appendix}
\usepackage[ruled,vlined]{algorithm2e} 
\usepackage[belowskip= 2pt,aboveskip=-2pt,labelfont=bf]{caption}
\usepackage{lineno}
\usepackage{setspace}
\usepackage{xcolor}
\usepackage{multirow}

\begin{document}

\begin{frontmatter}

\title{An algorithm for non-parametric estimation in state-space models}

\author[ren]{Thi Tuyet Trang Chau\corref{cor1} \fnref{TC}}%
\ead{trang.chau@lsce.ipsl.fr}
\author[bre]{Pierre Ailliot} 
\ead{pierre.ailliot@univ-brest.fr}
\author[ren]{Val\'erie Monbet} 
\ead{valerie.monbet@univ-rennes1.fr}

\cortext[cor1]{Corresponding author}
\fntext[TC]{Present address: LSCE, CEA Saclay, 91191 Gif-sur-Yvette cedex, France}

\address[ren]{IRMAR-INRIA, University of Rennes, Rennes, France}
\address[bre]{Univ Brest, CNRS, LMBA - UMR 6205, Brest, France}

\begin{abstract}

State-space models are ubiquitous in the statistical literature since they provide a flexible and interpretable framework for analyzing many time series. In most practical applications, the state-space model is specified through a parametric model.
However, the specification of such a parametric model may require an important modeling effort or may lead to models which are not flexible enough to reproduce all the complexity of the phenomenon of interest. In such situation, an appealing alternative consists in inferring the state-space model directly from the data using a non-parametric framework. The recent developments of powerful simulation techniques have permitted to improve the statistical inference for parametric state-space models. It is proposed to combine two of these techniques, namely the Stochastic Expectation-Maximization (SEM) algorithm and Sequential Monte Carlo (SMC) approaches, for non-parametric estimation in state-space models. The performance of the proposed algorithm is assessed though simulations on toy models and an application to environmental data is discussed.

\end{abstract}

\begin{keyword}
 State-space models \sep Non-parametric statistics \sep SEM algorithm \sep Local linear regression \sep Conditional particle filter
\end{keyword}

\end{frontmatter}

\doublespacing

\section{Introduction}
\label{intro}

\textbf{State-space models (SSMs)} provide a natural framework to study time series with observational noise  in  environment, economy, computer sciences, etc. They have a wide range of applications in data assimilation, system identification, model control, change detection, missing-data imputation \citep[see e.g.][]{durbin2012time}. The general SSM which is considered in this paper is defined through the following equations,
\begin{numcases}{\label{eq: SSM}} 
X_{t} = m \left(X_{t-1},Z_t \right) + \eta_{t},  \quad   [hidden] \label{eq: evo-mod} 
\\
Y_t = H_t(X_t)  + \epsilon_t, \quad \quad  [observed] \label{eq: obs-mod}.
\end{numcases}

The dynamical model $m$ describes the time evolution of the latent process $\{X_t\}$. It may depend on some covariates (or control) denoted $\{Z_t\}$. The operator $H_t$ links the latent state to the observations $\{Y_t\}$. The random sequences $\{\eta_t\}$ and $\{\epsilon_t\}$ model respectively the random components in the dynamical model and the observational error. Throughout this paper, we make the classical assumptions that $H_t$ is known (typically $H_t(x)=x$) and that $\{\eta_t\}$ and $\{\epsilon_t\}$ are independent sequences of Gaussian distributions such that 
$\eta_{t} \stackrel{iid}{\sim}  \mathcal N \left(0,Q_t(\theta)\right)$ and $\epsilon_{t} \stackrel{iid}{\sim} \mathcal N \left(0,R_t(\theta)\right)$ where $\theta$ denotes the parameters involved in the parameterization of the covariance matrices. 

In this paper, we are interested in situations where the dynamical model $m$ is unknown or numerically intractable. To deal with this issue, a classical approach consists in using a simpler parametric model to replace $m$. However, it is generally difficult to find an appropriate parametric model which can reproduce all the complexity of the phenomenon of interest. In order to enhance the flexibility of the methodology and simplify the modeling procedure, non-parametric approaches have been proposed to estimate $m$. 

Such \textbf{non-parametric SSMs} were originally introduced in \citet{tandeo2015combining, lguensat2017analog} for data assimilation in oceanography and meteorology. In these application fields,  a huge amount of historical data sets recorded using remote and in-situ sensors or obtained through numerical simulations are now available and this promotes the development of data-driven approaches. It was proposed to build a non-parametric estimate $\widehat m$ of $m$  using the available observations and plug this non-parametric estimate  into usual filtering and smoothing algorithms to reconstruct the latent space  $X_{1:T}=(X_1,...,X_T)$ given observations $y_{1:T}=(y_1,...,y_T)$. Numerical experiments on toy models show that replacing $m$ by $\widehat m$  leads to similar results if the sample size used to estimate $m$ is large enough to ensure that $\widehat m$ is "close enough" to $m$. Some applications to real data are discussed in \citet{fablet2017data}.

Various non-parametric estimation methods have been considered to build surrogate nonlinear dynamical models in oceanography and meteorology. The more natural one is probably the nearest neighbors method known as the Nadaraya-Watson approach in statistics \citep{fan2008nonlinear} and analog methods in meteorology \citep{yiou2014anawege}. In \cite {lguensat2017analog}, better results were obtained with a slightly more sophisticated estimator known as local linear regression (LLR) in statistics  \citep[][]{cleveland1988locally} and constructed analogs in meteorology \citep{tippett2013constructed}.  More recently, it has been proposed to use other machine learning (ML) tools such as deep learning 
\citep[see][]{bocquet2019datapa} or sparse regression \citep[see][]{brunton2017chaos} to better handle high dimensional data and mimic the behaviour of numerical methods used to approximate the solutions of physical models. 

\begin{figure*}[ht!]
\centering
\includegraphics{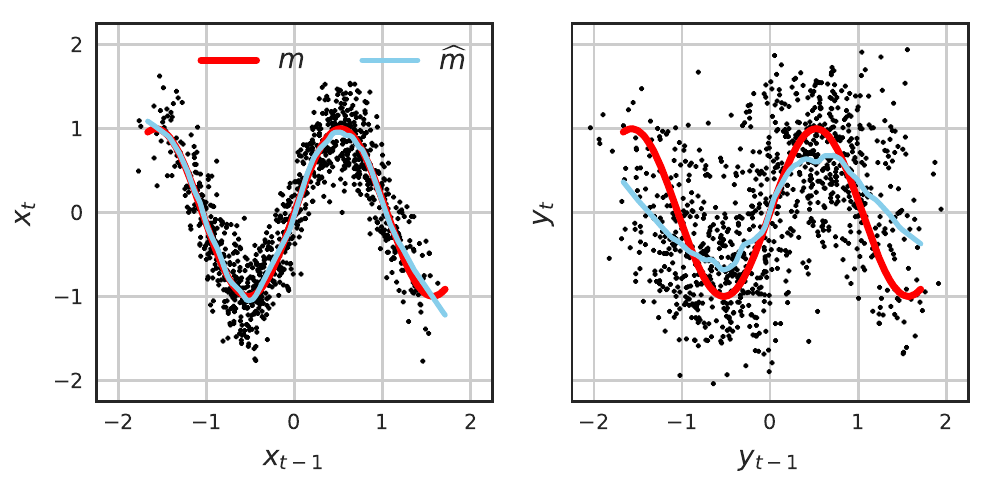} 
\caption{ Scatter plots of $\left(x_{t-1},x_{t}\right)$ (left) and $\left(y_{t-1},y_{t}\right)$ (right) for a sequence simulated with the SSM defined by (\ref{eq: sin}) and corresponding LLR estimate $\widehat m$ of $m$.}
\label{fig:prob-sin}
\vspace*{-0.5cm}
\end{figure*}

In the above mentioned references, it is generally assumed that a sequence $x_{1:T}=(x_1,...,x_T)$  of "perfect" observations with no observational error is available to estimate the dynamical model $m$. However, in practical applications, only a sequence $y_{1:T}$ of the process $\{Y_t\}$ with observational errors is given to fit the model. The main contribution of this paper is to propose a method to build non-parametric estimate of $m$ in this context. A simple approach would consist in "forgetting" the observational errors and computing directly a non-parametric estimate based on the sequence $y_{1:T}$ but this may lead to biased estimates. This is illustrated on Figure~{\ref{fig:prob-sin}}  obtained with the toy SSM defined as
\begin{align} \label{eq: sin}
\begin{cases}
X_{t} = \sin \left(3X_{t-1} \right) +  \eta_{t},  \quad \eta_{t} \stackrel{iid}{\sim} \mathcal{N} \left(0,Q \right)
\\
Y_t = X_t  +\epsilon_t, \quad \epsilon_t \stackrel{iid}{\sim} \mathcal{N} \left(0,R \right)
\end{cases}
\end{align}
with $Q=R=0.1$. The left plot shows the scatter plot $\left(x_{t-1},x_{t}\right)$ for a simulated sequence $(x_1,...,x_T)$ of the latent process $\{X_t\}$ and the corresponding non-parametric estimate $\widehat m$ based on this sample which is reasonably close to $m$. The right plot shows the scatter plot $\left(y_{t-1},y_{t}\right)$ of the corresponding sequence with observation noise. Note that $Y_t$ is obtained by adding a random noise to $X_t$ and this has the effect of blurring the scatter plot by moving the points both horizontally and vertically. It leads to a biased estimate of $m$ when the non-parametric estimation method is computed on this sequence. In a regression context, it is well known from the literature on errors-in-variables models that  observational errors in covariates lead, in most cases, to a bias towards zero of the estimator of the regression function \citep[see][]{carroll2006measurement}. One of the classical approaches to reduce the bias is to introduce instrumental variables which help to get information about the observational error. This approach has been adapted to linear first order auto-regressive models in \citet{meijer2013measurement} and further studied in \citet{lee2017many}. \citet{carroll2006measurement} gives an overview of different methods to build consistent estimators in the context of regression. Among them, we notice the local polynomial regression and the Bayesian method for non-parametric estimation but, as far as we know, they have not been generalized for time series.  

From a statistical point of view, the proposed model is semi-parametric with a parametric component for the white noise sequences whose distributions are described by a parameter $\theta$ and a non-parametric component for the dynamical model $m$. The Expectation-Maximization (EM) algorithm and its variants \citep[see e.g.][]{dempster1977maximum} are often used to fit parametric models with a latent component. The E-step of the EM algorithm consists in computing the smoothing distribution, i.e. the conditional distributions of the latent space given the observations. The smoothing distributions can generally not be computed analytically in nonlinear state-space models. However, the recent development of powerful simulation techniques, known as Sequential Monte Carlo (SMC) methods, permits to produce nowadays accurate approximations of the smoothing distributions at a reasonable computational cost and this allows to perform statistical inference in nonlinear state-space models \citep[see][for a recent review]{kantas2015particle}. In \citet{chau2018efficient}, it was proposed to use the Conditioning Particle Filter-Backward Simulation (CPF-BS) algorithm  presented for instance in 
\citet{ lindsten2013backward} in the E-step of the EM algorithm. It was found, through numerical experiments, that the combination of CPF-BS algorithm and EM recursions leads to an efficient numerical procedure to estimate the parameters of parametric SSMs. 

This paper discusses an extension of the algorithms proposed in \citet{chau2018efficient} to non-parametric SSMs where the non-parametric estimate of $m$ is updated at each iteration of the EM recursions using the  trajectories simulated from the smoothing algorithms. It permits to correct sequentially the bias in the estimate of $m$ due to observational errors. This method can also be interpreted as a generalization of the Bayesian approach of \citet{carroll2006measurement} for time series.

The  paper is organized as follows. The estimation of the parametric component using EM recursions is introduced in Section~\ref{sec: 2}. Then this algorithm is extended to estimate both the parametric and non-parametric components in Section~\ref{sec: 3}. Simulation results obtained on a toy model (Lorenz-63) are presented in Section~\ref{sec: 4}.  Then, Section~\ref{sec: 5} discusses an application to oceanographic data, where the algorithms are used to impute missing wave data given noisy observations. The paper ends with some concluding remarks in Section~\ref{sec: 6}. All the codes of the proposed approach used for numerical experiments in this paper are available on \url{https://github.com/tchau218/npSEM}.

\section{SEM algorithm for parametric estimation in SSMs} 
\label{sec: 2}

In this section, it is assumed that the dynamical model $m$ is known or that a surrogate model has already been fitted and the estimation of the unknown parameter  $\theta$, given a sequence $y_{1:T}$ of noisy observations, is discussed. The notation $\mathfrak{M}$ stands for the true dynamical model $m$ if it is known, or for the surrogate model otherwise. Remark that the covariate $Z_t$ which appears in \eqref{eq: evo-mod} is omitted in Sections~\ref{sec: 2}-\ref{sec: 4} for the sake of simplification.

The EM algorithm is probably the most usual algorithm to perform maximum likelihood estimation in models with latent variables including SSMs. It is an iterative algorithm where, at each iteration $r\geq 1$, the parameter value $\widehat \theta_{r-1}$ is updated through  the following steps.
\begin{itemize}
\item \textbf{E-step}: compute the smoothing distribution $p\left(x_{0:T}|y_{1:T};\widehat \theta_{r-1}\right)$ defined as the conditional distribution of the latent sequence $X_{0:T}$ given the sequence of observations $y_{1:T}$ and the current parameter value $\widehat \theta_{r-1}$.
\item \textbf{M-step}:  maximize the intermediate function 
\begin{align} 
I\left(\theta|\widehat \theta_{r-1},\mathfrak{M}\right)
\triangleq \int & \log p \left(x_{0:T},y_{1:T}; \theta,\mathfrak{M} \right) p(x_{0:T}|y_{1:T}; \widehat \theta_{r-1}) ~\mathrm{d}x_{0:T}
\label{eq: ExpL}
\end{align} 
obtained by integrating the complete log-likelihood function 
\begin{align} 
\label{eq:full}
\log p \left(x_{0:T},y_{1:T}; \theta,\mathfrak{M} \right)=  & \log p\left(x_0 \right)+ \sum_{t=1}^T \log p\left(x_t|x_{t-1};Q_t(\theta),\mathfrak{M}\right)  \\
\nonumber
  & + \sum_{t=1}^T \log p\left(y_t|x_t;R_t(\theta)\right)
\end{align} 
over the smoothing distribution computed in the E-step where the initial distribution $ p\left(x_0 \right)$ is assumed to be known. $p\left(x_t|x_{t-1};Q_t(\theta),\mathfrak{M}\right)$ and $p\left(y_t|x_t;R_t(\theta)\right)$ denote respectively the transition kernel of the Markov process $\{X_t\}$ defined by \eqref{eq: evo-mod} and the conditional probability distribution function of $Y_t$ given $X_t=x_t$ associated to \eqref{eq: obs-mod}. 

Finally, the parameter value is updated as 
\begin{equation}
    \label{eq:M}
    \widehat \theta_r=\arg\max \limits_{\theta } I\left(\theta|\widehat \theta_{r-1},\mathfrak{M}\right).
\end{equation}
\end{itemize}

For nonlinear (non-Gaussian) SSMs, the smoothing distributions do not have any tractable analytical expression. However, sequential Monte Carlo (SMC) algorithms \citep[see][for instance]{cappe2007overview, godsill2004monte}
 allow to generate sequences of these conditional distributions. 
They provide  samples (particles) $\{\tilde x_{0:T,r}^{(i)}\}_{i = 1:N}$   to approximate the smoothing distribution  $p(.|y_{1:T}; \widehat \theta_{r-1})$ with the empirical distribution
\begin{equation} \label{eq: emp-smo-dis}
\widehat p_r (\mathrm{d}x_{0:T}|y_{1:T}) = \frac 1 N \sum_{i=1}^{N} \delta_{\tilde x_{0:T,r}^{(i)}}(\mathrm{d}x_{0:T}),
\end{equation}
where $\delta_{x}$ denotes the point mass at $x$. Replacing the true smoothing distribution by this empirical distribution in \eqref{eq: ExpL} to estimate the intermediate function $I$   of the EM algorithm leads to the so-called Stochastic EM (SEM) algorithm. 

One of the key points of SEM algorithms is to compute an approximation of the intermediate function $I$ at a reasonable computational cost. If the number of particles $N$ is large, then the law of large numbers implies that
$$
\widehat I\left(\theta|\widehat \theta_{r-1},\mathfrak{M}\right)
\triangleq \frac 1 N \sum_{i=1}^{N}  \log p \left(\tilde x_{0:T,r}^{(i)},y_{1:T}; \theta, \mathfrak{M} \right) $$
is a good approximation of $I\left(\theta|\widehat\theta_{r-1},\mathfrak{M}\right)$ and the SEM algorithm is close to the EM  algorithm. Various extensions of the SEM algorithm have been proposed to reduce  the size of simulated samples $N$ and save computational time \citep{wei1990monte, delyon1999convergence}. The SEM algorithms and their variants using particle filters \citep[see][]{kantas2015particle} suffer from another issue. In order to get samples which approximately follow the smoothing distribution a large amount of particles is typically required. Since the smoothing algorithm has to be run at each iteration of the EM algorithm, this may lead to prohibitive computational costs  \citep[see][for a recent review]{fearnhead2018particle}.

Kalman-based algorithms, such as the Ensemble Kalman Smoother (EnKS), are traditionally used in the data assimilation community since they generally provide good approximations of the smoothing distributions with a low number of particles  \cite[see][]{carrassi2018data}. However, they are based on Gaussian approximations which may not be suitable for nonlinear SSMs. 

Conditional SMC (CSMC) samplers, which are based on combinations of SMC and Markov Chain Monte Carlo (MCMC) approaches, have been developed as alternatives to particle filters and Kalman-based algorithms. The first CSMC samplers, called Conditional Particle Filters (CPFs), have been introduced by \citet{andrieu2010particle, lindsten2012ancestor}
 and were combined with the EM algorithm in \citet{lindsten2013backward}. The CPF algorithms simulate samples of $x_{0:T}$ conditionally on the current value of the parameter ${\theta}$ and a trajectory in the latent state space referred to as the conditioning sequence. The conditioning sequence is updated sequentially and this builds a Markov chain which has the exact smoothing distribution $p(\mathrm{d}x_{0:T}|y_{1:T};{\theta})$ as an invariant distribution \citep[see][ for numerical illustrations]{svensson2015nonlinear, chau2018efficient}.

Nevertheless, as many sequential smoothing algorithms, when the length $T$ of the observed sequence is large, the CPF algorithms suffer from "sample impoverishment" with all the trajectories sharing the same ancestors. A way  to reduce impoverishment is to run a Backward Simulation  (BS) algorithm after the CPF one. Backward simulation, proposed initially in \citet{godsill2004monte}, is a natural technique to simulate the smoothing distribution given the (forward) filter outputs \citep[see][]{lindsten2013backward}. This leads to the Conditional Particle Filter-Backward Simulation (CPF-BS) sampler (see Algorithm~\ref{alg: 3} in Appendix). Recently, \citet{chau2018efficient} proposed to use the CPF-BS smoothing algorithm in conjunction with the SEM algorithm (see Algorithm~\ref{alg: 1} below). The authors found experimentally that the method  outperforms several existing EM algorithms in terms of both state reconstruction and parameter estimation, using low computational resources.  

\begin{algorithm}
{\bf  Initialization}: choose an initial parameter value $ \widehat \theta_{0}$ and an initial conditioning sequence $x_{0:T}^{*}$. \\
For $r \geq 1$,\\
\textbf{(1) E-step}: generate $N$ trajectories $\{\tilde x_{0:T,r}^{(i)}\}_{i = 1:N}$ of the smoothing distribution using the CPF-BS algorithm (\ref{alg: 3}) with parameter value $\widehat{\theta}_{r-1}$, dynamical model $\mathfrak{M}$, conditioning sequence $x_{0:T}^{*}=\tilde x_{0:T,r-1}^{(1)}$, and observations $y_{1:T}$,
\\
  \textbf{(2) M-step}: update the parameter value 
  $$\widehat{\theta}_r = \arg\max \limits_{\theta }~ \widehat I\left(\theta|\widehat \theta_{r-1},\mathfrak{M}\right),$$
 end.
 \caption{SEM algorithm for parametric SSMs [SEM($\mathfrak{M}$)]}
 \label{alg: 1}
 \end{algorithm}

\section{Non-parametric estimation in SSMs} 
\label{sec: 3}

In the previous section, it is assumed that the true dynamical model $m$ is known or that a surrogate model is available but this may be unrealistic for some applications. In this section, the joint estimation of $\theta$ and non-parametric estimation of $m$  from a sequence $y_{1:T}$ with observational error is discussed.

Following the numerical results presented in \cite{lguensat2017analog}, Local Linear Regression (LLR) is used to build a non-parametric estimate of $m$. The idea of LLR is to locally approximate $m$ by a first-order Taylor's expansion, $m \left(x'\right) \approx m(x)+\nabla m(x) (x'-x)$, for any $x'$ in a neighborhood of $x$. In practice, the intercept $m(x)$ and the slope $\nabla m(x)$  are estimated by minimizing a weighted mean square error where the weights are defined using a kernel. In this study the tricube kernel is used as in \citet{cleveland1988locally}. This kernel has a compact support and is smooth at its boundary. 
Throughout the paper, LLR is performed based on the $k$-nearest neighborhood of $x$ \citep[][]{altman1992introduction}. In this case, the support of the kernel is defined as the smallest rectangular area which contains the $k$ nearest neighbors and the kernel bandwidth adapts to the density of points in the neighborhood of $x$.

As mentioned in the introduction, applying the LLR estimation method on a sequence $y_{1:T}$ with observational errors leads to a biased estimate for $m$ (see Figure~\ref{fig:prob-sin}). In order to reduce sequentially the bias induced by the observation noise, it is proposed in Algorithm~\ref{alg: 2}
to update the non-parametric estimate of $m$ at each iteration of the SEM algorithm. Algorithm~\ref{alg: 2} is similar to Algorithm~\ref{alg: 1}, with the exception that the non-parametric estimate $\widehat{m}_{r}$ of the dynamical model is updated at each iteration $r$ of the EM algorithm using LLR on the trajectories of the smoothing distribution simulated in the E-step. This estimate is then used in the smoothing algorithm and in the complete log-likelihood function (\ref{eq:full}) which appears in the definition of the intermediate function of the EM algorithm.

\begin{algorithm}
{\bf  Initialization}: choose an initial parameter value $ \widehat \theta_{0}$, an estimate of the dynamical model $\widehat{m}_{0}$, and  a conditioning sequence $x_{0:T}^{*}$. \\
{For $r \geq 1$},\\
\textbf{(1) E-step}: generate $N$ trajectories $\{\tilde x_{0:T,r}^{(i)}\}_{i = 1:N}$ using a sequential Monte Carlo smoothing algorithm with parameter value $\widehat{\theta}_{r-1}$, dynamical model $\widehat{m}_{r-1}$, conditioning sequence $x_{0:T}^{*} = \tilde x_{0:T,r-1}^{(1)}$ and observations $y_{1:T}$,
\\
  \textbf{(2) M-step}: 
  \begin{enumerate}
  \item[i.] Parameter update: compute the parameter value 
  $$\widehat{\theta}_r = \arg\max  \limits_{\theta } ~\widehat I\left(\theta|\widehat \theta_{r-1},\widehat{m}_{r-1} \right),$$
  \item[ii.] Catalog update: compute an LLR estimate $\widehat{m}_r$ of $m$ based on the 'updated catalog' $\left\{\tilde x_{0:T,r}^{(i)} \right\}_{i=1:N}$, \\
 \end{enumerate}
  end.
 \caption{{SEM-like algorithm for non-parametric SSMs [npSEM]}} 
 \label{alg: 2}
 \end{algorithm}

\begin{figure}[ht]
\centering
\includegraphics[scale=0.9]{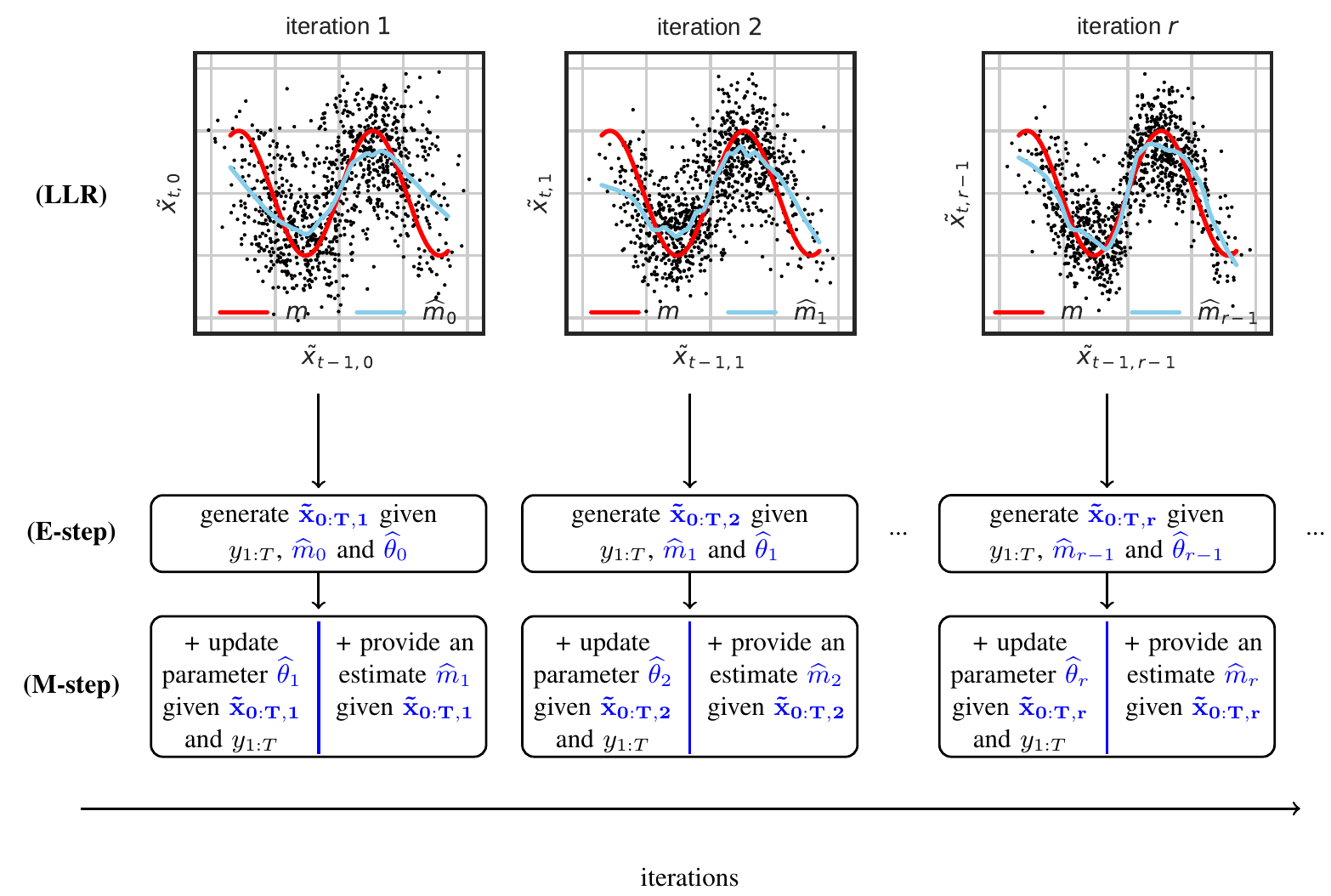}
\caption{An illustration of Algorithm~\ref{alg: 2} (npSEM) on the sinus SSM \eqref{eq: sin}. } 
\label{fig:npSEM-illus-sin}
\end{figure}

The name "SEM-like" algorithm is used to highlight that the proposed algorithm shares similarities with the SEM algorithm for parametric estimation in SSMs. Remark that it is not an SEM algorithm because the M-step is not only composed of a likelihood maximization for $\theta$, as in the usual EM algorithm, but also of a 'catalog update' for $m$. This is illustrated on Figure~\ref{fig:npSEM-illus-sin} using  the sinus model \eqref{eq: sin}. At each iteration, the estimate $(\widehat m_r)_{r \geq 0}$ of the dynamical model $m$ is updated using LLR on the sample of the smoothing distribution simulated in the E-step. It allows to correct sequentially the bias in the non-parametric estimate of $m$ which is induced by the observation error. Such EM-like and SEM-like algorithms have been proposed in the literature for fitting mixture models  \citep{zhang2002algorithm, benaglia2009like} with non-parametric estimates of the distributions in the different components of the mixture. The spirit of Algorithm~\ref{alg: 2} is also close to the one of the iterative global/local estimation (IGLE) algorithm of \citet{young2010mixtures} for estimation of mixture models with mixing proportions depending on covariates. 

Remark that the smoothing sample $\{\tilde x_{t,r-1}^{(i)}\}_{i \in \{1,...,N\}}$ at time $t$ depends on the  observation $y_t$ at the same time, and that over-fitting may occur if this sample is used to build the non-parametric estimate $\widehat{m}_{r-1}$ and propagate the particles at time $t$ in the smoothing algorithm at iteration $r$ of the EM algorithm. This over-fitting was confirmed using numerical experiments.
To tackle this issue, at each iteration $r$ and for each time $t$, $\widehat{m}_r (\tilde x_{t-1,r}^{(i)})$ is estimated using LLR based on the subsamples  $\{ \tilde x_{0:(t-\ell) \cup (t+\ell):T,r-1}^{(i)}\}_{i \in \{1,...,N\}}$ where the smoothing sequences $\{ \tilde x_{(t-\ell+1):(t+\ell-1),r-1}^{(i)}\}_{i \in \{1,...,N\}}$ are removed from the learning sequence. The lag $\ell$ is chosen as a priori such that the correlation between $Y_{t-\ell}$ and $Y_{t}$ is low. At each iteration, the LLR estimate of $m$ is updated and the number $k$ of nearest neighbours needs to be chosen. Cross-validation technique is used to select an optimal value of $k$.

The numerical complexity of Algorithm~\ref{alg: 2} is mainly linked to the nearest neighbor search which has to be performed for the LLR estimation.  At each iteration $r$, the nearest neighbor search is repeated for each discrete time $t \in \{1,\cdots,T\}$ and for each particle  $i \in \{1,\cdots,N\}$, thus $NT$ searches are performed. Furthermore, when the catalog is updated the cross-validation needs to be performed to update the optimal number of neighbors and this adds more nearest neighbor searches. The nearest neighbor search is carried out by using  the {\it introselect} algorithm \citep{musser1997introspective} which has a complexity of $O(M)$ in the best cases and $O(M log M)$ in the worst cases. Here, $M$ denotes the size of the learning data set. For the univariate SSM (\ref{eq: sin}), $56.5$ minutes of CPU time are necessary to run $100$ iterations of the npSEM algorithm ($N = 10, T=1000$) using Python on a computer with a 3-GHz CPU and 128-GB of RAM.

\section{Simulation results} 
\label{sec: 4}

In the previous sections, the sinus SSM \eqref{eq: sin} was used as an illustrative example. Many simulation experiments were performed using this model and the obtained results were generally satisfactory. Some results are reported in a Supplementary Material document  for the sake of brevity. This section focuses on the more challenging 3-dimensional Lorenz model \citep[see in][]{lorenz1963deterministic} which is one of the favorite toy models in data assimilation since it is  a sophisticated (nonlinear, non-periodic, chaotic) but low-dimensional dynamical system 
\citep{lguensat2017analog, bocquet2019datapa}.
The considered Lorenz-63 (L63)  SSM  on $\mathbb{R}^3$ is defined as  
\begin{equation} \label{eq: L63-SSM}
\begin{cases}
X_t =  m(X_{t-1}) +  \eta_t, \quad \eta_t \stackrel{iid}{\sim} \mathcal{N} \left( 0, Q\right)\\
Y_t =   X_t +  \epsilon _t, \quad   \epsilon _t \stackrel{iid}{\sim} \mathcal{N} \left( 0, R\right).
\end{cases}
\end{equation}

Covariance matrices in the above model are assumed to be diagonal and proportional to the identity matrix $I_3$ of dimension $3$, such that $Q=\sigma _Q^{2}I_3$ and $R=\sigma _{R}^{2} I_{3}$ with true parameter values $\sigma _Q^{2}=1$ and $\sigma _{R}^{2}=4$. The dynamical model $m$ at any value $x$ in $\mathbb{R}^3$ is computed by integrating the following differential system
\begin{equation} \label{eq: L63-ODE}
\begin{cases}
z(0) = x \\
\frac{dz(\tau)}{d\tau} = g(z(\tau)), \quad   \tau  \in [0,\mathrm{d}t] \\
m(x) = z(\mathrm{d}t)
\end{cases}
\end{equation}
where $g(z)  =\left(10(z_2-z_1), ~ z_1(28-z_3) -z_2,~  z_1z_2 - 8/3 z_3 \right)$, $~ \forall z=(z_1,z_2,z_3) \in \mathbb{R}^3$. For each time $t$, the system of  ordinary differential equations  \eqref{eq: L63-ODE} is integrated by running a Runge-Kutta scheme (order 5). The value  of $\mathrm{d}t$ is fixed to 0.08 which corresponds to a 6-hour time step in the observation of  atmospheric data and  was considered in the works of \citet{dreano2017estimating, lguensat2017analog}.

Given an observed sequence, the eight algorithms listed below are run and compared.
\begin{itemize}
    \item 'CPF-BS update' corresponds to the npSEM algorithm \ref{alg: 2}.
		\item 'CPF-BS no update' corresponds to the npSEM algorithm~\ref{alg: 2} where the 'catalog update' step in not performed. In this algorithm, only the value of the parameter $\theta$ is updated at each iteration  but the non-parametric estimate of $m$ is not updated. Hence it also corresponds to  the SEM($\mathfrak{M}$) algorithm \ref{alg: 1} with $\mathfrak{M}$ obtained using LLR on the observed sequence $y_{1:T}$. 
		\item 'CPF-BS perfect' corresponds to  the SEM($\mathfrak{M}$) algorithm~\ref{alg: 1} with $\mathfrak{M}$ obtained using LLR on a perfect sequence. Remark that in order to run this algorithm, a realization of the true state $x_{1:T}$ ('perfect catalog') needs to be available, which is generally not the case for real applications. 
		\item 'CPF-BS true $m$' corresponds to  the SEM($\mathfrak{M}$) algorithm~\ref{alg: 1} with $\mathfrak{M}=m$. In this algorithm, the true dynamical model is assumed to be known.
		\item 'EnKS update', 'EnKS no update', 'EnKS perfect' and 'EnKS true $m$' are the same algorithms as defined above except that the smoother used in the E-step is the EnKS instead of the CPF-BS.
\end{itemize}

Hereafter, the parametric estimations obtained from the eight algorithms are compared. Then the ability of the algorithms to reconstruct the state is evaluated with respect to the length of the learning time series and with respect to $dt$ which is related to the strength of the non-linearities in the observed sequence. 
Unless stated otherwise, observed sequences $y_{1:T}$ of length $T=1000$ are simulated and $150$ iterations of the different algorithms are run. In order to evaluate the variability of the estimators,  each experiment is repeated on $30$ independent sequences. The number of members in the EnKS algorithm is set equal to $20$. This is a classical value used in the literature \citep{dreano2017estimating}. The number of particles $N_f$ of the conditional particle filter is fixed equal to $N_f=10$ and the number of realizations $N_s$ for the backward simulation step is fixed equal to $N_s=5$  (see Appendix). These values have been chosen empirically. The lag $l=5$ is used in the npSEM algorithms. Inspired by the application, the  EM algorithm coupled with a Kalman Smoother is run to initialize the algorithms (see Section~\ref{sec: 5} for more details). In practice, $100$ iterations of this algorithm is performed. Then, the estimate of $\theta$ is set as the initial parameter value $ \widehat \theta_{0}$ of all the eight algorithms, and the mean of the smoothing distribution is taken as the initial conditioning sequence $x_{0:T}^{*}$ in the algorithms combined with the CPF-BS. The initial estimate $\widehat m_0$ of $m$ for the 'EnKS update' and 'CPF-BS update' algorithms is also obtained by using LLR on the mean of the smoothing distribution derived from the EM algorithm.

\begin{figure*}[ht]
\centering
\hspace*{-0.5cm}
\includegraphics[scale=.52]{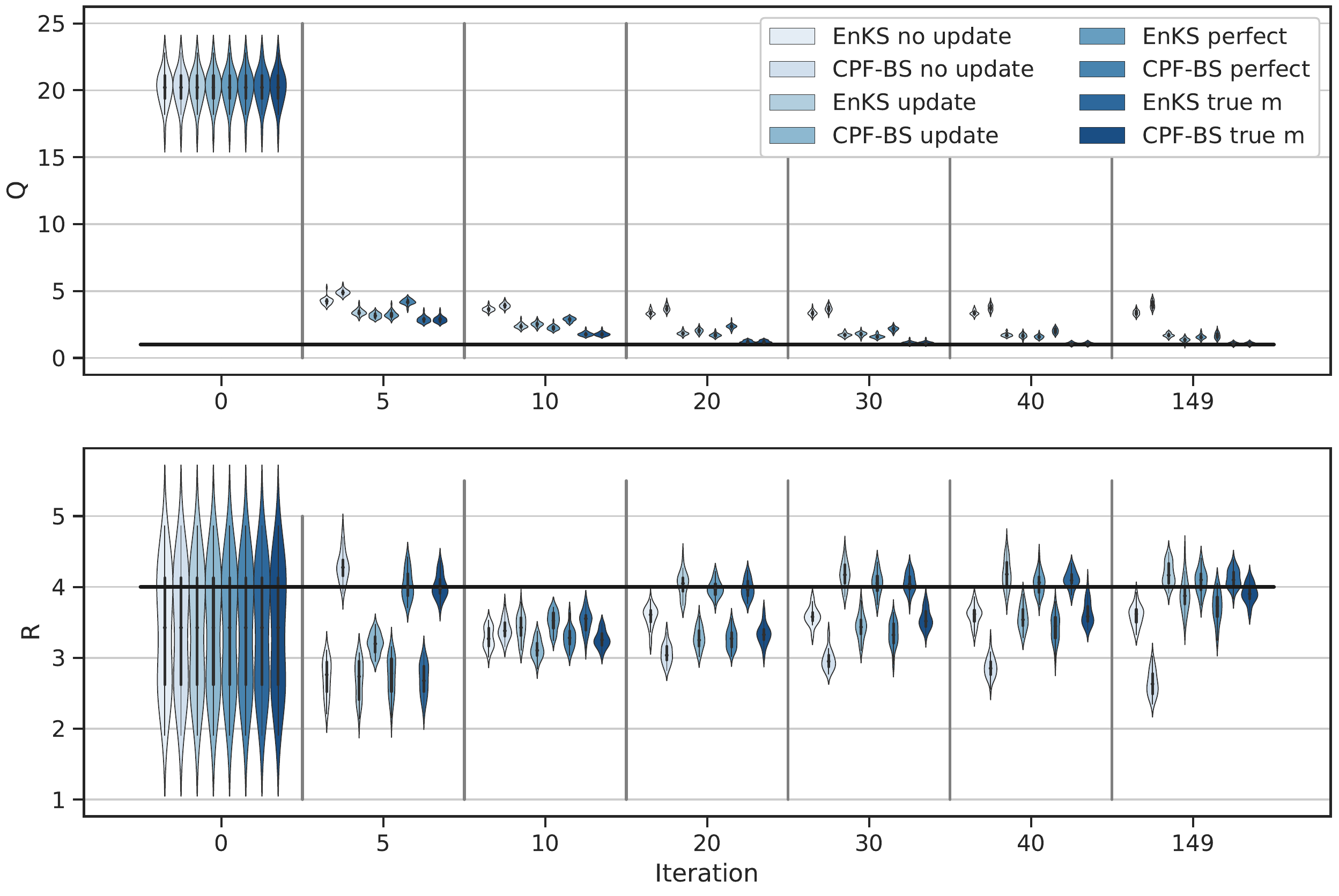} \\
	\caption{Distribution of the estimation in $Q$ and $R$ with respect to the iteration number for the L63 model~\eqref{eq: L63-SSM}. The horizontal black lines materialize the true values. The vertical gray lines separate the iteration blocks. In each block the performance of the eight algorithms is displayed. Results are obtained with $T=1000$.
	}
	\label{fig:conver-L63-R}
\end{figure*}

On Figure~\ref{fig:conver-L63-R}, the distributions of the estimates of the parameters $\sigma _Q^2$ and $\sigma _R^2$, which correspond to the diagonal coefficients of the covariance matrices $Q$ and $R$, are represented with respect to the iteration number using violin plots.  In general, the biases and variances of the estimates obtained using the eight algorithms have been significantly reduced after a few iterations. This is expected since the initial parameters are obtained using the  EM algorithm coupled with a Kalman Smoother which approximates the L63 model by a linear Gaussian model. The 'EnKS no update' and 'CPF-BS no update' algorithms  (1st and 2nd violin boxes of each block of 8) lead to the worst estimates. These algorithms based on a noisy version of the true state use a poor surrogate model in the forecast steps. In the 'EnKS update' and 'CPF-BS update' algorithms (3rd and 4th violin boxes), the catalog used to estimate the dynamic is updated at each iteration. It permits to iteratively reduce the observation errors in the catalog  and reduce the estimation error. They hence provide estimates close to the ones obtained with the 'EnKS perfect', the 'CPF-BS perfect' and the two SEM algorithms using the true L63 model. The CPF-BS algorithm is expected to better capture the non-linearities in the model compared to the EnKS algorithm and it is thus not surprising that the 'CPF-BS update' algorithm  slightly outperforms the 'EnKS update' algorithm.
\begin{figure*}[ht]
	\centering
	\hspace*{-0.5cm}
	\includegraphics[scale=.52]{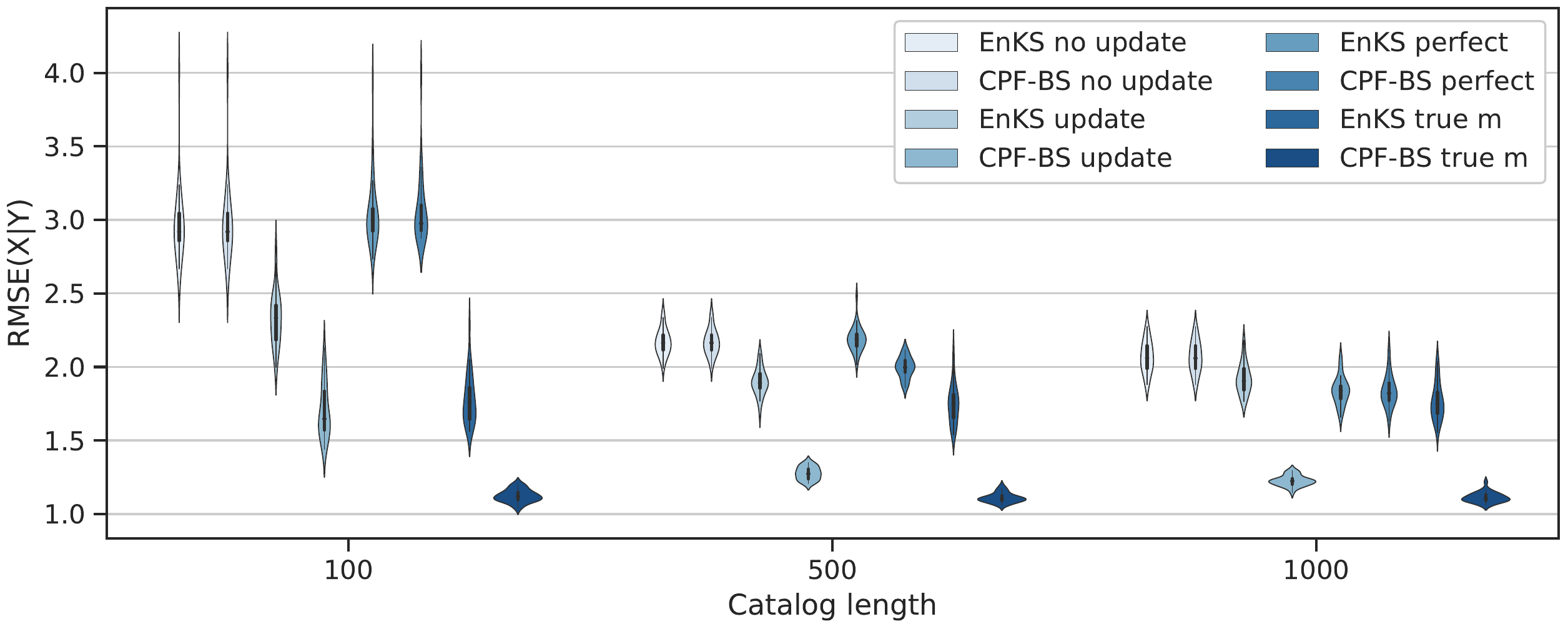}
     \includegraphics[scale=.52]{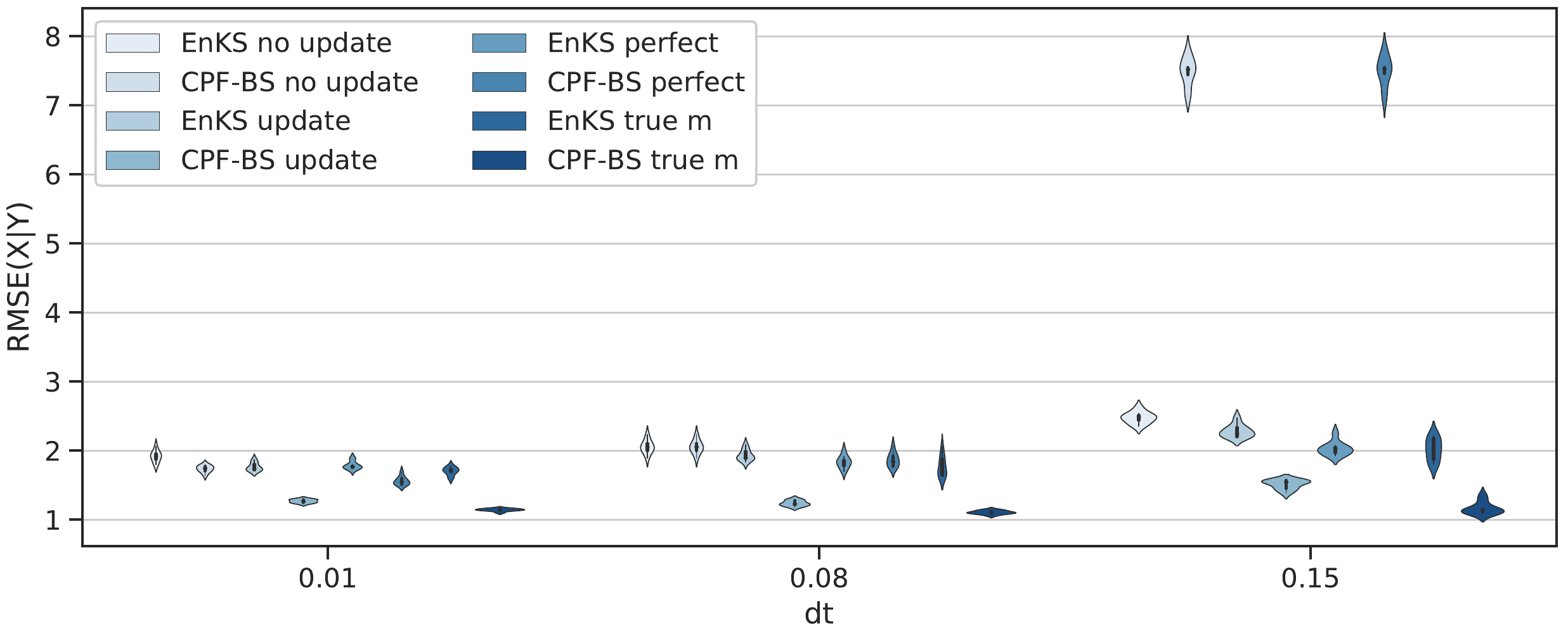}
	\caption{Reconstruction errors (RMSE$(X|Y)$) of the SEM and npSEM algorithms with respect to  different catalog length ($T$) and non-linearity degree ($\mathrm{d}t$) for the L63 model~\eqref{eq: L63-SSM}. }
	\label{fig:L63-RMSE}
\end{figure*}

In order to measure the global performance of the methodology, the reconstruction error whose computation is described hereafter is considered. First, validation sequences of $x_{1:T^ \prime }$ and  $y_{1:T^ \prime }$ ( $T^ \prime = 1000$) are generated with the true model. These validation sequences are independent from the learning time series.
At each iteration of the SEM and npSEM algorithms, the associated smoothing algorithm is run on the validation sequence $y_{1:T^ \prime }$ using the current estimates of $m$, $Q$ and $R$. Then, the sample mean $\bar x_t$ of the smoothed particles over the last  $10$  EM iterations is computed as an estimate of the conditional expectation of the latent state given the sequence of observations
 $E[X_t|y_{1:T^ \prime }]$. Finally, the Root of Mean Square Error (RMSE), 
\begin{equation} \label{eq: RMSE_smo}
\mathrm{RMSE}\left(X |Y \right)=  \sqrt{\frac{1}{T^ \prime}\sum\limits_{t=1}^{T^ \prime } (\bar x_t-x_t)^2 },
\end{equation} 
is used to assess the reconstruction skill of the algorithms.

The distributions of the reconstruction errors $RMSE(X|Y)$ for the eight algorithms are shown on Figure~\ref{fig:L63-RMSE} for different lengths $T \in \{100, 500, 1000\}$ of the learning sequences. As expected, the reconstruction errors decrease when $T$ increases and the algorithms do not seem to suffer too much from a degradation of the mixing properties of the smoothing algorithms. Again, the algorithms with catalog updates clearly outperform the algorithms with no update and the algorithms based on the CPF-BS algorithms outperform the ones based on the EnKS algorithms. It is also noteworthy that the algorithms with catalog updates clearly outperform the algorithms which use the perfect catalog $x_{1:T}$, and lead to results close to those obtained with the true model $m$ despite being calibrated using only  a noisy sequence $y_{1:T}$.
In order to investigate the robustness of the proposed methodology to highly nonlinear dynamics, the bottom panel of Figure~\ref{fig:L63-RMSE} shows the reconstruction error as a function of $\mathrm{d}t \in \{0.01,0.08,0.15\}$. Nonlinearities increase with $\mathrm{d}t$ and thus it is not surprising that the reconstruction errors generally increase with $\mathrm{d}t$. Remark however that the 'CPF-BS update' algorithm again performs well even when the nonlinearities are strong.

\begin{figure*}[ht]
\centering
\includegraphics[]{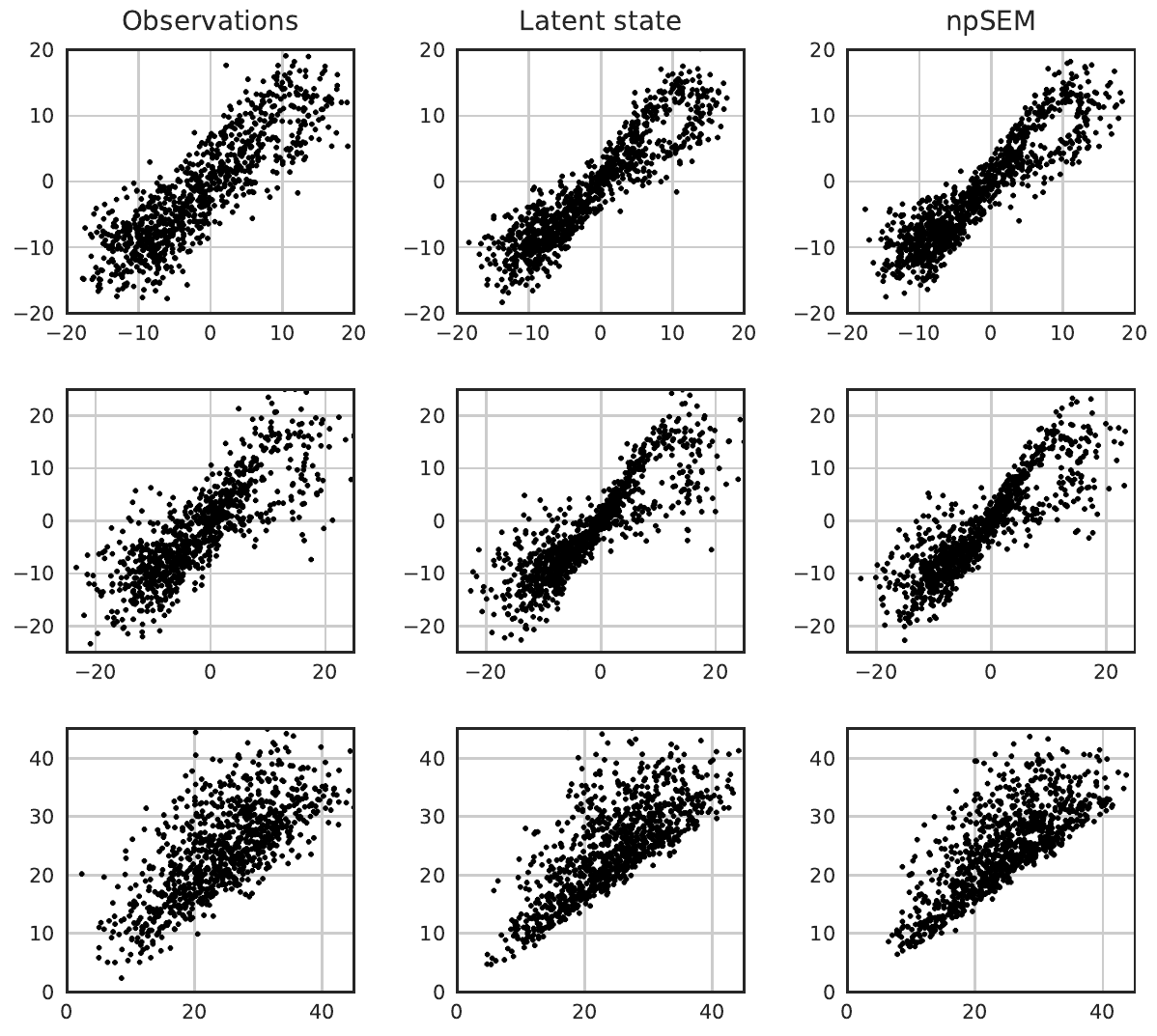}
\caption{Scatter-plots of $\left(y_{t-1},y_{t}\right)$ (left), $\left(x_{t-1},x_{t}\right)$ (middle) and $\left(\widetilde x_{t-1},\widetilde x_{t}\right)$ (right) for each of the three components of the L63 model defined by \eqref{eq: L63-SSM}. $\{\widetilde x_t\}$ stands for one of realizations generated at the final iteration of the 'CPF-BS update' algorithm.}
\label{fig:scatt-L63}
\end{figure*}

\begin{figure*}[ht]
\centering
\includegraphics[scale=.5]{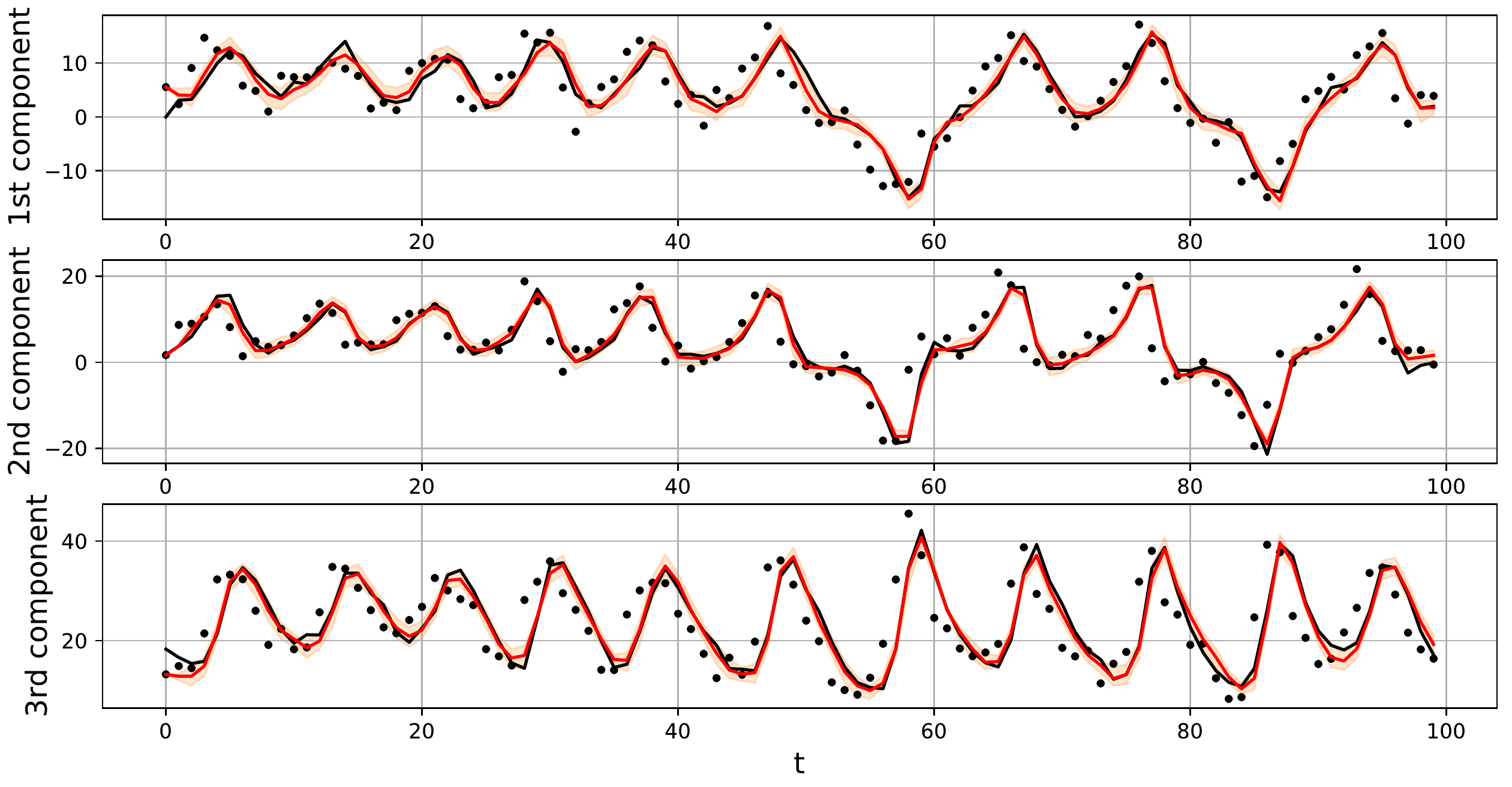}
\caption {
Time series of L63 components (black lines) with noisy observations (black dots) and smoothing distribution of the 'CPF-BS update' algorithm.
Empirical means (red lines) and $95 \%$ prediction intervals (red areas) are computed using the smoothing trajectories generated in the last 10 iterations of the 'CPF-BS update' algorithm. On this sequence the coverage probability is approximately $85\%$ for each component. 
}
\label{fig:mean-CIs-L63}
\end{figure*}

Figure~\ref{fig:scatt-L63} illustrates the  ability of the 'CPF-BS update' algorithm (\ref{alg: 2}) to reconstruct the dynamics of the three components of the L63 model. From left to right, the scatter plots correspond to successive values at time $t-1$ and $t$ of the observed sequence, the true state and a realization simulated at the last iteration of the algorithm. As for the sinus model used in the introduction, the comparison of the left and middle panel shows that the observation noise significantly blurs the true dynamic. The proposed algorithm efficiently reduces the noise and the scatter plots corresponding to the catalog in the last iteration of the 'CPF-BS update' algorithm (right panel) look very similar to the ones of the true dynamic (middle panel). It suggests that proposed methodology is successful in estimating the true dynamical model $m$. This could be useful in applications where this model is of interest and, for example, be used to build or validate a surrogate parametric model. 
This is also confirmed by the time series displayed on Figure~\ref{fig:mean-CIs-L63} which shows that the true state $x_t$ is generally close to the smoothing mean $\bar x _t$ and generally lies in the $95\%$ prediction interval.

\section{Real case study}
\label{sec: 5}

\begin{figure*}[ht]
    \centering
     \includegraphics[scale=.3]{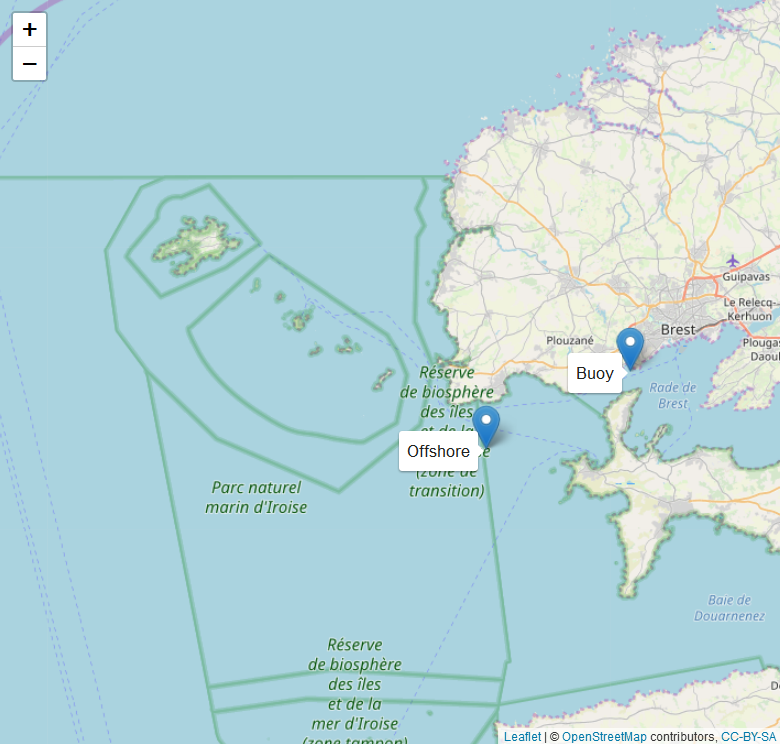}\\
    \hspace*{-1cm}
    \includegraphics[scale=.35]{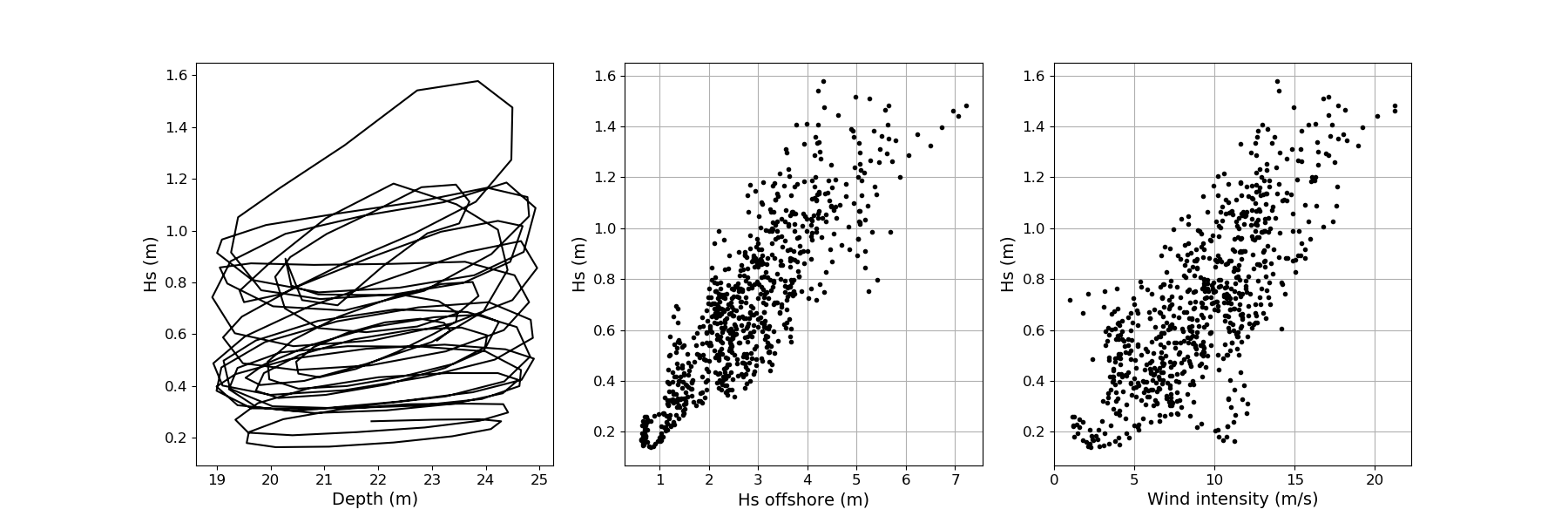}
    \caption{Top: map with buoy and offshore location indicated. Bottom:
    bivariate time series of $(H_s,D)$, (left panel), scatter plots of $(H_{s}^{(off)},H_s)$ (middle panel) and  $(H_s,U)$ in January 2016, at St Anne du Porzic (France).}
    \label{fig:map&Hs_depth}
\end{figure*}
Data imputation is a recurrent problem in many fields 
and particularly in oceanography since measurements are generally complicated to make over the ocean \citep{fablet2017data}. In this section, the proposed methodology is illustrated using data from the HOMERE data set. HOMERE is an hourly hindcast data set based on the WAVEWATCH III model (version 4.11) on an unstructured spatial grid covering the English Channel and Bay of Biscay over the period 1994-2016  \citep{boudiere2013suitable}. Our particular interest is the  significant wave height $H_s$. It is usually defined as 4 times the standard deviation of the surface elevation and was intended to mathematically express the height estimated by a "trained observer". We focus on the point with geographical coordinate $(48.3563^\circ N, 4.5508^\circ W)$ located in the entrance of the bay of Brest (see top panel of Figure~\ref{fig:map&Hs_depth}). At this location, where a buoy is located, wave conditions are influenced by the water depth $D$ which depends on the tide, local wind speed $U$ which creates wind sea and offshore wave conditions since the waves generally propagate to the east from the Atlantic Ocean inside the bay. Hereafter  $H_s^{(off)}$ denotes the significant wave height at the location shown on Figure~\ref{fig:map&Hs_depth} and $Z_t = (D_t,U_t,H_{s,t}^{(off)})\in \mathbb R^3$ the values of the covariates at time $t$. At the bottom of Figure~\ref{fig:map&Hs_depth}, the pairwise relations between $H_s$ and its covariates are shown. The tide induces cycles where $H_s$ grows with the depth and the link between $H_s$ and $H_s^{(off)}$  is linear in mean but exhibits heteroscedasticity. 

\begin{figure*}[ht]
    \centering
     \includegraphics[scale=.9]{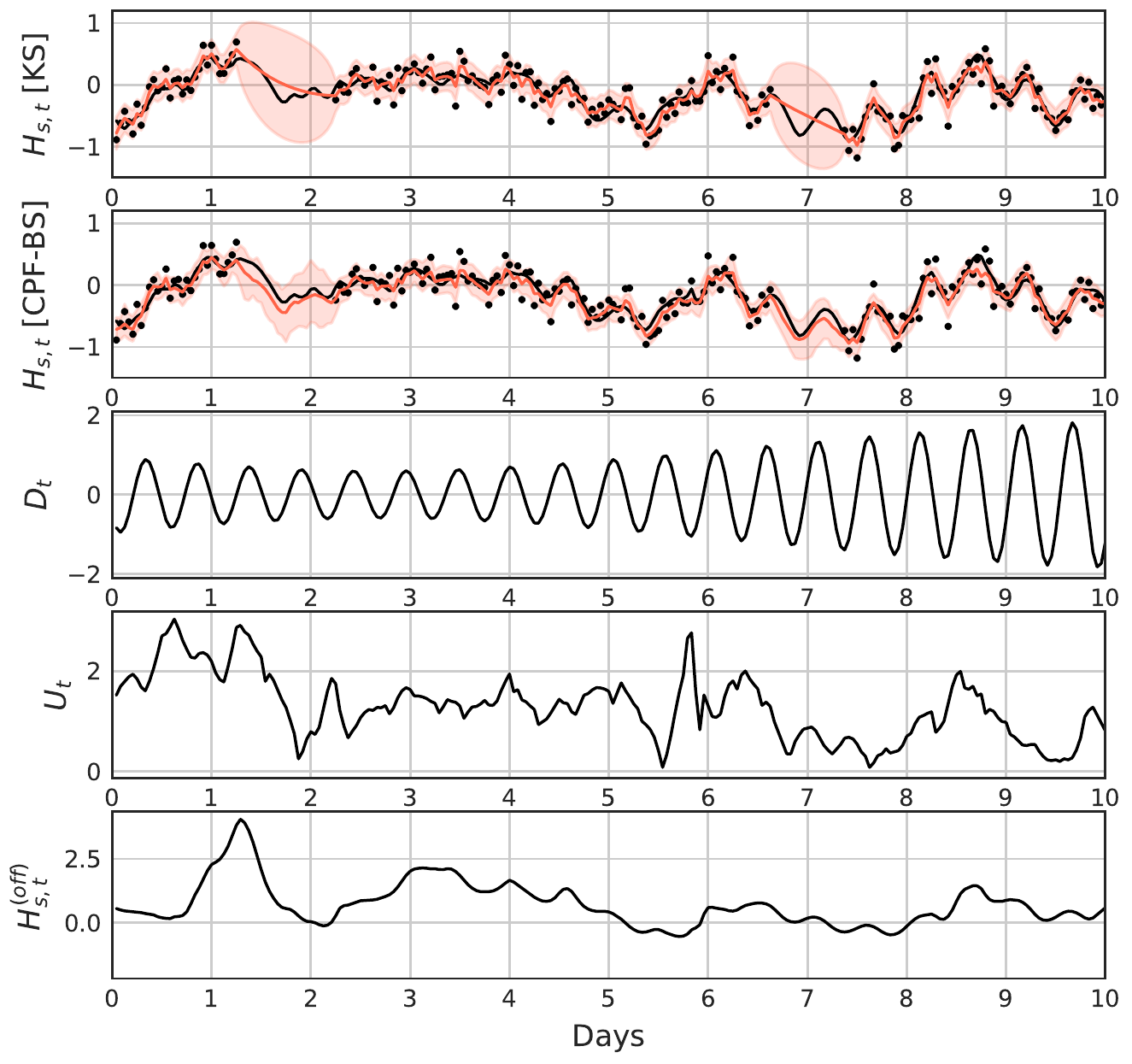}
  \caption{Log transformed $H_s$ (2 top panels) and covariates (other panels) in January 2016 at St Anne du Porzic (France). The observations with noise ${R}=0.04$
  are represented by the dots. The reconstruction of the state is represented by the red plain line for the linear SSM (1st panel) and 'CPF-BS update' algorithm (2nd panel).  The shaded areas represents a 95\% prediction interval. }
    \label{fig:Hs_covariates}
\end{figure*}

In the numerical experiments, the $H_s$ from the hindcast data set at the buoy location is supposed to be the true state which we wish to reconstruct. As usual in the literature, a logarithm transform is applied to $H_{s,t}$ \citep{o1984weather} and the result is denoted $X_t$. In order to mimic the behaviour of noisy observations recorded at the buoy location, an artificial time series $\{Y_t\}$ is simulated by adding a white noise to $\{X_t\}$ according to the observation equation (\ref{eq: obs-mod}). Different levels of observation noise $\sqrt{R} \in \{0.1, 0.2, 0.5\}$ are considered since it might impact the performance of the reconstruction. Some gaps corresponding to missing data are also created in the time series $\{Y_t\}$. On Figure \ref{fig:Hs_covariates}, the time series $\{X_t\}$ (plain lines), $\{Y_t\}$ (dots)  and $\{Z_t\}$ (bottom plots)  during $10$ days in January 2016 are shown with missing values around the 2nd and the 7th of January. Our goal is to impute the missing values as well as reconstruct the time series $\{X_t\}$ from the observed sequence $\{Y_t\}$ and the covariate sequence $\{Z_t\}$. The considered meteorological time series are non-stationary with an important seasonal and eventually inter-annual components. A pre-processing step can be applied to the data in order to remove these components. In this work, the inter-annual components (related e.g. to climate change) are neglected and seasonal components can be taken into account by fitting different models separately for the 12 calendar months. Here, we focus only on the January month.

We consider the state-space model (\ref{eq: SSM}) whose structure is summarized by the directed graph below. 
\begin{equation} 
				\nonumber
				\begin{array}[t]{lccccccccc}
 			\text{Covariates}	& \cdots & \rightarrow & Z_{t-1} &  \rightarrow & 	Z_{t} & \rightarrow & Z_{t+1} & \rightarrow & \cdots \\
 				& & (m,Q) & \downarrow & & \downarrow & & \downarrow  & & \\
 			\text{Latent state}	& \cdots & \rightarrow & X_{t-1} &  \rightarrow & 	X_{t} & \rightarrow & X_{t+1} & \rightarrow & \cdots \\
 				& & (H,R) & \downarrow & & \downarrow & & \downarrow  & & \\
\text{Observations
} & \cdots &  & Y_{t-1} &  & Y_{t} & & Y_{t+1} &  & \cdots \\
				\end{array}
				\end{equation}
			
\begin{figure}[ht]
    \centering
     \includegraphics[scale = .8]{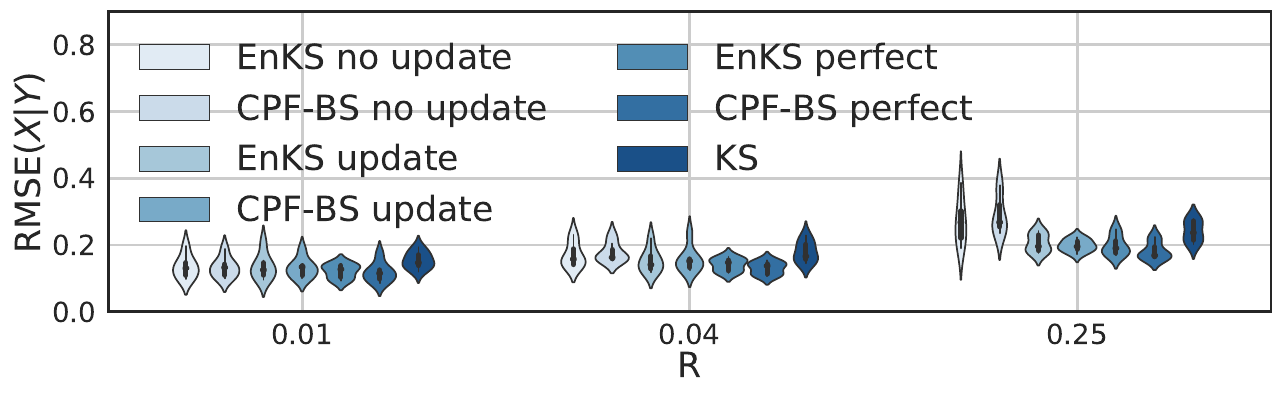} \\
     \includegraphics[scale = .8]{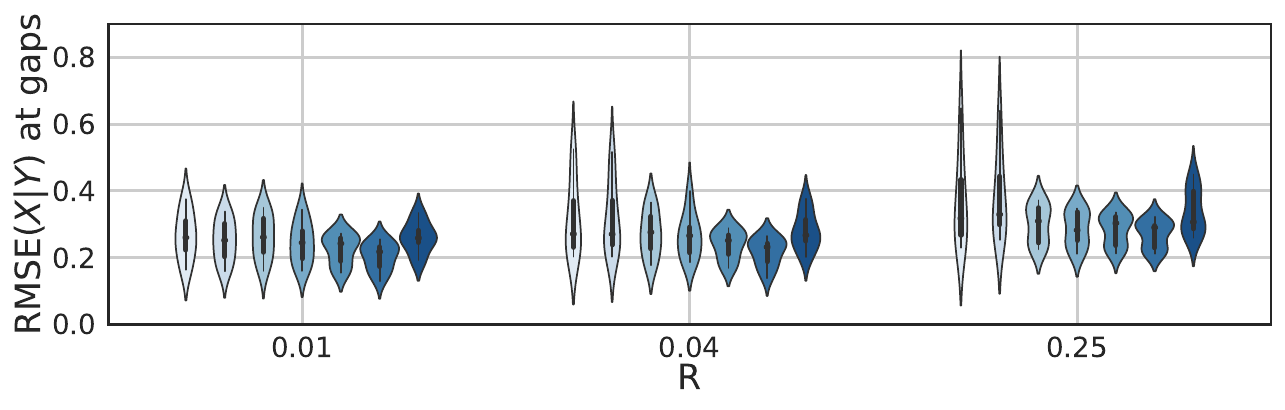} 
	\caption{Reconstruction errors computed on the validation data of log transformed $H_s$ in January 2016 at St Anne du Porzic (France) with respect to  different values of observation variance $R$. Top: $RMSE(X|Y)$ computed between  the reconstructed time series and the true state, bottom: $RMSE(X|Y)$ computed at the time steps where data are missing (gaps). Each violin box is obtained from results of $10$ repetitions of each algorithm run on different learning sequences, which are randomly sampled in the January months of the period 1994-2015.}
    \label{fig:Hs_RMSE}
\end{figure}

The dynamical operator $m$ is  unknown but it can be estimated using the non-parametric methodology introduced above where the covariates are used in the nearest neighbor search step. In order to evaluate the global performance of the algorithm, a reconstruction error is computed as in the previous section. More precisely, the time series is split into two parts where the first one (data in January for the period 1994-2015) is used to fit the model and the second one (data in January of 2016) is used as a validation sequence to compute the reconstruction error (\ref{eq: RMSE_smo}). In the experiments, the performances of the algorithms introduced in the previous sections are compared with the ones obtained with a linear Gaussian state-space model, which is defined by (\ref{eq: SSM}) where $m \left(X_{t-1},Z_t \right) = \alpha X_{t-1} +\beta$ with $\alpha$ and $\beta$ real parameters, and $H(x)=x$. The unknown parameters are estimated using the  EM algorithm combined with the Kalman smoother (KS) which provides analytic and exact solutions for the smoothing distributions. This approach  is one of the most usual approach to model time series with observation noise.
	
Figure~\ref{fig:Hs_RMSE} shows the reconstruction errors $RMSE(X|Y)$. On the top panel, the error is estimated using the whole  time series (reconstruction error)
whereas on the bottom plot the error is computed only based on the time step where the data are missing (imputation error).  Let us first notice that the algorithms based on CPF-BS only slightly outperform the ones based on EnKS. This may be a sign that, conditionally to the covariates, the non-linearities in the dynamics of $H_s$ dynamic are not strong.  As expected, the reconstruction errors and the imputation errors tend to increase when the variance of the noise increases and the catalog update helps to reduce the reconstruction error, especially when the variance of the observation noise is large. 

The imputation error is slightly stronger than the reconstruction error. It is expected because the values are missing over quite large time intervals (about 24h). In the middle of these gaps, there is less information on $H_s$ and the reconstruction error increases. It is clearly visible on  Figure \ref{fig:Hs_covariates} which shows a times series of $10$ days of $H_s$ as well as those of the covariates. On the two  plots, the dots are the observations. The red line and shaded areas materialize respectively the empirical mean and a 95\% prediction interval of the smoothing distributions. It shows that the npSEM algorithm clearly outperforms the results obtained with the EM algorithm coupled with Kalman smoother in terms of bias and variance. The introduction of covariates in the npSEM algorithm allows to reproduce the tide cycle and include the information brought by the wind and offshore wind conditions even in long gaps with missing data when no information on the $H_s$ condition at the buoy location is available.

\section{Conclusions and perspectives}
\label{sec: 6}

This paper introduces an npSEM algorithm for non-parametric estimation in SSMs. Numerical experiments on toy models and oceanographic data show that it permits to successfully reconstruct the latent state space from noisy observations and estimate the latent dynamic $m$.

The proposed methodology has only been validated on low dimensional time series and more works have to be done in order to handle higher dimensional problems since both the particle filters and the nearest-neighbors estimation methods suffer from the curse of dimensionality. Combining recent advances in the particle filters for higher dimensional systems \citet[see e.g.][]{beskos2017stable} with advanced machine learning approaches for estimating $m$ \citep[see e.g.][]{bocquet2019datapa,fablet2017bilinear} may allow to tackle higher dimensional problems with the additional advantage of leading to a reduction of computational costs if the machine learning tool is efficiently implemented.

The proposed approach is based on the assumption that the noise sequences are Gaussian, which may be restrictive for practical applications. In order to give preliminary ideas  about the performance of the algorithms when the model is miss-specified some additional numerical experiments are detailed in the Supplementary Material. Data are generated from an SSM with Student noises and the algorithms of the paper are used to reconstruct the state time series. It shows that the reconstruction error slightly increases when the tails of the noise are heavier than the ones of a Gaussian distribution, but the algorithms seem to be pretty robust.

\section*{Appendix} \label{sec: App}
\begin{algorithm}
\noindent
{\small
\textbf{Inputs}: conditioning trajectory $ X^* = x_{0:T}^*$, observations  $y_{1:T}$ and fixed parameter $\theta$. 
\begin{enumerate}
\item[\textbf{1}.] Run \textbf{CPF} algorithm with the inputs given to obtain a system of $N_f$ particles and their weights $(x_t^{(i)},w_t^{(i)})_{t =0:T}^{i=1:N_f}$.
\begin{itemize}
\item Initialization:
	\begin{itemize}
	\item[+] Sample $\{x_0^{(i)} \}_{i =1:N_f} \sim ~ p_\theta(x_0)$ and set $x_0^{(N_f)} = x_0^*$.
	\item[+] Set initial weights $w _0^{(i)} =1/N_f, \forall{i =1:N_f} $.
	\end{itemize}
\item For $t =1:T$,
	\begin{itemize}
    \item[+] {Resample} indices $\{I_t^{i}\}_{i =1:N_f}$ of potential particles with respect to the previous weights $(w _{t-1}^{(i)})_{i=1:N_f}$.
    \item[+]  Propagate new particle 
      \begin{align} 
      \nonumber
	  x_t^{(i)} \sim  p_\theta \left(x_t|x_{t-1}^{(I_{t}^i)} \right), \forall{i =1:N_f}.	
      \end{align}
    \item[+]  Replace for the conditioning particle, $x_t^{(N_f)} = x_t^*$ and $I_t^{N_f} =N_f$.
    \item[+]  Compute the weight 
    \begin{align}
    \nonumber
    w_{t}^{(i)} =\frac{ p_\theta(y_t|x_t^{(i)})}{\sum \limits _{i = 1}^{N_f}  p_\theta(y_t|x_t^{(i)})}, \forall {i =1:N_f} 
    \end{align}
   \end{itemize}
   end for.
\end{itemize}
\item[\textbf{2}.] Repeat the following \textbf{BS} algorithm using the outputs of the \textbf{CPF} algorithm to gets $N_s$ trajectories $\{\tilde x_{0:T}^j\}_{j=1:N_s}$.
\begin{itemize}
  \item For $t = T$, draw $\tilde x_T^j$ following the discrete distribution $p(\tilde x_T^j = x_T^{(i)}) = w_T^{(i)}$.
   \item For $t <T$,\\
  + Calculate smoothing weights 
  \begin{align} 
  \nonumber
  \tilde{w}_t^{(i)} {=}  \frac{p_\theta(\tilde x_{t+1}^{j}| x_{t}^{(i)})~ w_{t}^{(i)}}{\sum \limits_{j=1}^{N_f} p_\theta(\tilde x_{t+1}^{j}| x_{t}^{(i)})~ w_{t}^{(i)}}, ~ \forall i =1:N_f. 
  \end{align}
  + Draw  $\tilde x_{t}^j$ with respect to $p (\tilde x_{t}^j = x_t^{(i)}) = \tilde{w}_{t}^{(i)}$.\\
  end for
  \end{itemize} 
\item[\textbf{3}.] Update the  new conditioning trajectory $X^*$  by sampling uniformly from $N_s$ trajectories.
\end{enumerate}
\textbf{Outputs}: realizations describing the smoothing distribution $p_\theta(x_{0:T}|y_{1:T})$.
}
\caption{Smoothing with Conditional Particle Filter-Backward Simulation (CPF-BS)}
\label{alg: 3}
\end{algorithm}


\bibliographystyle{plainnat}
\pagebreak

\end{document}